\documentclass[sigconf, nonacm]{acmart}

\usepackage{enumitem}

\newcommand{\MSD}[2]{$M=#1, SD=#2$}
\newcommand{\mishal}[2]{\mis{}: #1, \hal{}: #2}
\newcommand{\selfcgpt}[2]{\self{}: #1, \cgpt{}: #2}

\newcommand{\squote}[1]{`#1'}
\newcommand{\dquote}[1]{``#1''}

\newcommand{\itemD}[1]{\item[\textbf{\textit{D#1:}}]}
\newcommand{\itemQ}[1]{\item[\textbf{\textit{RQ#1:}}]}

\newcommand{\self}{\textit{Self Ideation}}
\newcommand{\cgpt}{\textit{Co-GPT Ideation}}
\newcommand{\rank}{\textit{Ranking}}

\newcommand{\mis}{\textsc{Misinformation}}
\newcommand{\hal}{\textsc{Hallucination}}

\AtBeginDocument{
  \providecommand\BibTeX{{
    \normalfont B\kern-0.5em{\scshape i\kern-0.25em b}\kern-0.8em\TeX}}}

\begin{document}

\title{Rapid AIdeation: Generating Ideas With the Self and in Collaboration With Large Language Models}

\author{Gionnieve Lim}
\email{gionnievelim@gmail.com}
\orcid{0000-0002-8399-1633}
\affiliation{
  \institution{Singapore University of Technology and Design}
  \streetaddress{8 Somapah Rd}
  \country{Singapore}
  \postcode{487372}
}

\author{Simon T. Perrault}
\email{perrault.simon@gmail.com}
\orcid{0000-0002-3105-9350}
\affiliation{
  \institution{Singapore University of Technology and Design}
  \streetaddress{8 Somapah Rd}
  \country{Singapore}
  \postcode{487372}
}

\renewcommand{\shortauthors}{Lim and Perrault}

\begin{abstract}
Generative artificial intelligence (GenAI) can rapidly produce large and diverse volumes of content. This lends to it a quality of creativity which can be empowering in the early stages of design. In seeking to understand how creative ways to address practical issues can be conceived between humans and GenAI, we conducted a rapid ideation workshop with 21 participants where they used a large language model (LLM) to brainstorm potential solutions and evaluate them. We found that the LLM produced a greater variety of ideas that were of high quality, though not necessarily of higher quality than human-generated ideas. Participants typically prompted in a straightforward manner with concise instructions. We also observed two collaborative dynamics with the LLM fulfilling a \textit{consulting} role or an \textit{assisting} role depending on the goals of the users. Notably, we observed an atypical \textit{anti-collaboration} dynamic where participants used an antagonistic approach to prompt the LLM.
\end{abstract}



\maketitle

\section{Introduction}

With the rapid progress of GenAI in recent years, capabilities that were previously thought to be capable only by humans such as art and creativity are now achievable by technology to similar or even better standards. As such, there has been a rising interest in understanding the potentials and shortfalls of these technologies and in what capacities they can best serve users. Research in human-AI interaction and computer supported cooperative work has purported for such technologies to serve complementary and collaborative roles.

The topic of GenAI and creativity has garnered strong research interest due to the alignment of generative technology with creative tasks. In the creative process, one aspect in which GenAI may flourish is in the brainstorming stage where the goal is to produce as many novel ideas as possible within a span of time. Research has found promising results in the use of GenAI in the ideation process~\cite{tholander2023designideation, wan2023prewriting}.

Extending these studies, we consider a specific type of brainstorming, rapid ideation, which is a popular technique used with design teams. We conducted a workshop where participants ($N=21$) were asked to first ideate solutions to a given issue by themselves, and then to use ChatGPT to help them generate ideas. From content analysis of the ideas and prompts, we found that ideating with the LLM led to more unique ideas that were of high quality, yet they were not necessarily better than human-generated ideas. Prompts were typically written in a straightforward manner to convey concise requests using a semi-formal tone. We observed two forms of human-LLM collaboration where LLMs fulfilled a \textit{consulting} role or an \textit{assisting} role depending on the intentions of the users. Of note was a third \textit{anti-collaboration} dynamic, observed in rare instances, where participants tried to threaten or interrogate the LLM to get their desired outcomes instead.

In the remainder of the paper, we describe related work, the method and results, and discuss various conclusions and considerations from our findings. Through this preliminary study, we seek to extend existing knowledge on human-AI collaboration in creative tasks by contributing to the understanding of the capabilities of LLMs in rapid ideation tasks and by adding nuance to the collaborative dynamics between humans and AI.

\begin{figure*}[!htb]
  \centering
  \includegraphics[width=.9\linewidth]{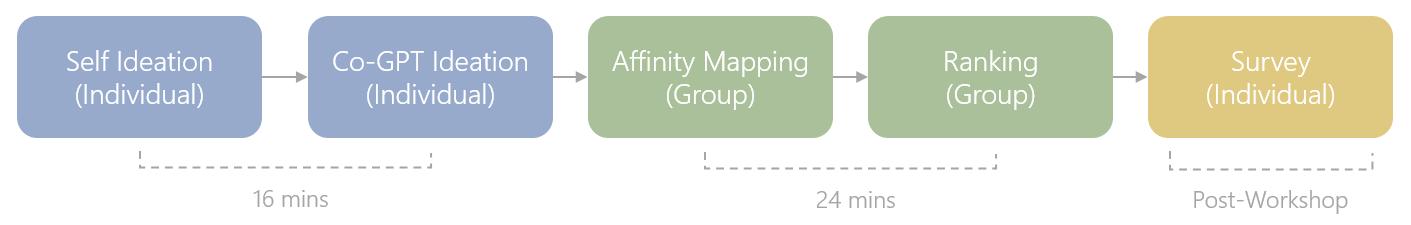}
  \caption{Phases of the workshop}
  \label{fig:workshop}
\end{figure*}

\section{Related Work}

Our work is situated among studies that have explored the use of GenAI in the creative process. With a focus on brainstorming, we seek to understand how LLMs enhance the co-creation process and how collaboration takes place.

\subsection{Brainstorming and the Creative Process}

Brainstorming was introduced by Osborn~\cite{osborn1957appliedimagination} in 1957 to improve the emergence of ideas across several phases. The initial phase includes defining the goal and organizing the brainstorming session. The next phase is the generation of ideas where focus is placed on quantity and participants are encouraged to have unconventional and innovative ideas and to combine and improve them, free from criticism. The last phase involves consolidating, evaluating and identifying the most relevant ideas. This technique has been polished over the decades with numerous research studies identifying methodological refinements~\cite{rossiter1994brainstorming}.

The purpose of brainstorming is to produce creative ideas - ideas that are original and useful~\cite{amabile1996creativity}. This involves the creative process, conceptualized by Guilford~\cite{guilford1967humanintel} in 1967 to employ two information processing modes: divergent thinking and convergent thinking. Divergent thinking is the unstructured and unbounded exploration of the design space for a task. Convergent thinking is synthesis and resolution of ideas to identify the best ones that fit the constraints of a task. These correspond to the ideation and evaluation phases in brainstorming respectively.

A popular brainstorming technique, particularly for design teams, is rapid ideation\footnote{\url{https://www.mural.co/blog/rapid-ideation}}, which centers on producing a large number of ideas in a short amount of time. While this technique can be intensive for humans, it essentially describes what GenAI is capable of, thus we look at how GenAI, when involved in brainstorming, can contribute to the creative process.

\subsection{Co-Creation and Collaboration with LLMs}

The quality of LLM productions have drawn considerable interest in its potential applications~\cite{sohail2023chatgpt}. With strong generative capacities, LLMs are well aligned with creative purposes, and there is much research interest on their capabilities. LLMs have been adopted in many creative tasks, meeting with various levels of success. While they fall short of transformational creativity, LLMs are capable of generating creative content of value, novelty and surprise~\cite{franceschelli2023creativity}. In studies comparing ideas produced by humans and LLMs, AI-generated ideas were observed to be better than human-generated ideas~\cite{girotra2023llmideas, joosten2024compareideation}. LLMs are also capable of generating more creative ideas through self evaluation~\cite{summers-stay2023glmcreativity}.

Delving into co-creation, studies have examined how LLMs can be adapted to enhance the creative process. In a workshop where expert participants used commercial GenAI tools to develop design concepts~\cite{tholander2023designideation}, participants noted that the advantages of GenAI included helping to save time and quickly mapping out the design space. However, they were concerned about the innovativeness of the ideas as the GenAI lacked contextual understanding of the problem. As the GenAI produced ideas rapidly, they also perceived the GenAI to cover the breadth more than the depth of the design space. In another study where participants used a commercial LLM for prewriting~\cite{wan2023prewriting}, participants used the LLM for coming up with ideas, organizing their thoughts and producing the desired text. While participants typically held a dominant role in the co-creation process, having their own preference for the writing and using the LLM to embellish them, they also considered ideas generated by the LLM, and even iteratively built upon those ideas, in the initial stages. Notably, the LLM helped some participants recover from writer's block, underscoring the collaborative dynamic between humans and LLMs.

While in line with the above studies, our work extends them in two ways. First, we offer insights on how GenAI supports the \textit{rapid} ideation process. Given that rapid ideation is more time-pressed, participants would have to be more efficient, which we posit will influence their prompting of the LLM and selection of ideas. Second, we provide prompts written by participants, discussing their prompting style and behavior when using the LLM to generate ideas.

\section{Method}

To understand how GenAI may contribute to brainstorming and the creative process, we investigated this through a design workshop centered on rapid ideation.
The choice of data collection through a workshop was to provide a more engaging and dynamic environment for ideating.
Workshops have also been used in similar work~\cite{tholander2023designideation, ekvall2023chatgptux}. The study was approved by an Institutional Review Board.

\begin{figure*}[!htb]
  \centering
  \includegraphics[width=\linewidth]{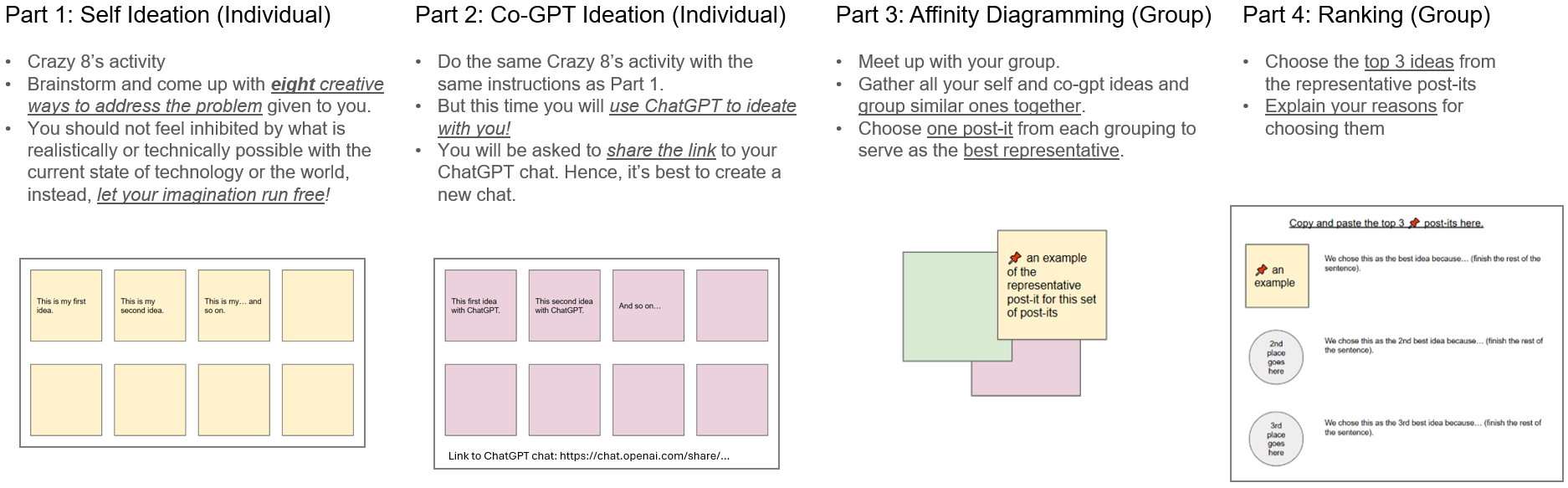}
  \caption{Instructions for the workshop activities}
  \label{fig:sheets}
\end{figure*}

\subsection{Workshop}

The workshop was conducted online using a video conferencing tool. Participants were briefed on the format of the workshop and given a document containing instructions and a link to the slides with the activity sheets. Figure~\ref{fig:workshop} shows the phases of the workshop which ran for 40 minutes which was based on the protocol for rapid ideation. After the briefing, participants broke into random groups of three and proceeded with the activities. The workshop organizer mainly acted as timekeeper yet remained available to answer any questions.

Each group was assigned to one of two issues where they had to come up with \dquote{ways to utilize ChatGPT to address online misinformation} (\mis{}) and \dquote{ways to address the hallucination issue of ChatGPT} (\hal{}). These issues were defined as \dquote{false content that is spread online} and \dquote{when ChatGPT gives an inaccurate answer to the question} respectively. These issues were chosen because they are key issues related to LLMs, one from humans' misuse of them and the other stemming internally from the technology\footnote{These issues have been covered by both specialist and mainstream news media globally.}.

The instructions and activity sheets were drafted to be self-explanatory (see Figure~\ref{fig:sheets}). Participants first completed two individual ideation activities based on the Crazy 8's process~\cite{knapp2016sprint} where they rapidly brainstormed ideas to address their group's given issue. This was done by entering their ideas onto \squote{post-its} in their own slides\footnote{The post-its were colored distinctly such that it was possible to distinguish which participant created the post-it and whether it was from the Self Ideation or Co-GPT Ideation activity.}. The \self{} activity required them to think of ideas by themselves whereas the \cgpt{} activity required them to use ChatGPT to generate ideas\footnote{All participants were previously experienced with using ChatGPT.}. Participants could optionally provide a link to their ChatGPT conversation history.

Next were the group activities. In the \textit{Affinity Diagramming} activity based on the KJ method~\cite{kawakita1967affinitymap}, the group members pulled together their individual post-its onto a single slide and grouped ideas that were similar. They also had to \squote{pin} the post-it that best represented their respective cluster of post-its using a pushpin emoji. They then proceeded to identify the top three ideas from the representative post-its during the \rank{} activity and explain their choices. No explicit ranking metric was provided to the participants, they discussed freely and internally within their group, and wrote the reasons for their ranking decisions in the activity sheets thereafter.

A survey link was made available after the workshop for participants to optionally complete in their own time that asked for their demographic information and experience with ChatGPT.

\subsection{Participants}
The study involved 21 participants from the student population at a university. Participation was voluntary with no compensation provided. In the optional demographic survey submitted by 15 participants, 9 participants identified as women and 6 as men. 11 participants are 18-25 years old and 4 are 26-35 years old. 6 participants are undergraduate students, 6 are master's students and 3 are doctoral students. The participants were thus young and highly educated. They were also familiar with OpenAI's ChatGPT\footnote{\url{https://chat.openai.com/}}, a popular LLM service, with 8 participants using it at least 3 days a week. All participants were also aware that content from ChatGPT may be inaccurate, and generally leaned towards being concerned about receiving inaccurate responses (\MSD{3.87}{0.74} out of 5). ChatGPT was chosen because of its wide availability and familiarity for the participants.

\subsection{Analysis}

We analyzed three sets of data collected from the workshop:

\begin{itemize}[leftmargin=1cm]
    \itemD{1} Post-its from the \self{} and \cgpt{} phases
    \itemD{2} Top three representative post-its from the \rank{} phase
    \itemD{3} ChatGPT prompts from the \cgpt{} phase
\end{itemize}

In doing so, we sought to answer the following research questions:

\begin{itemize}[leftmargin=1cm]
    \itemQ{1} Which ideation process produces more unique ideas?
    \itemQ{2} Which ideation process produces better quality ideas?
    \itemQ{3} How do users prompt ChatGPT to obtain their desired outcomes?
\end{itemize}

The data was analyzed using exploratory quantitative (for D1) and qualitative (for D3) content analysis~\cite{bengtsson2016contentanalysis}. The authors followed an inductive approach by familiarizing ourselves with the data, identifying initial codes in the data and labeling them, then searching for semantic (for D1) and thematic (for D3) clusters, i.e., reasonable groups of codes, which were defined and named. We used descriptive statistics for D2.

\section{Results}

There were 4 groups that worked on the issue of \mis{} and 3 groups for \hal{}. In total, participants produced 108 post-its during \self{} (\mishal{70}{38}) and 119 post-its during \cgpt{} (\mishal{70}{49}). There were 59 prompts (\mishal{36}{23}) from 17 ChatGPT conversations that were voluntarily submitted out of 21 participants. We report the results for each research question hereafter.

\subsection{RQ1: Which ideation process produces more unique ideas?}

To address RQ1, all the post-its were clustered based on the commonality of the ideas expressed in them. A unique idea was considered to be a proposed solution (either one or a group of post-its) which did not strongly overlap with other proposed solutions. Singular post-its that were not clustered were thus considered to be a unique idea on their own. Refer to Appendix Section~\ref{sec:resultsRQ1} for the full results.

We observed 23.6\% more unique ideas produced with \cgpt{} than with \self{} in general (\selfcgpt{38}{47}). The trend is reflected in the breakdowns for both the \mis{} (\selfcgpt{21}{26}) and \hal{} (\selfcgpt{17}{21}) issues.

To give a sense of the diversity of ideas produced, we briefly describe some ideas that were unique to each ideation phase. For \mis{}, some ideas from \self{} included using ChatGPT to highlight biased viewpoints and to have LLMs debate between themselves about a piece of content. In \cgpt{}, ideas were about using ChatGPT to assess the logical reasoning of content, summarizing and cross-referencing to other content, and directing to fact-checking websites. For \hal{}, ideas from \self{} involved having ChatGPT highlight generated content that might be problematic or to provide links to external information. In \cgpt{}, some ideas were to ensure that ChatGPT is trained on a good set of data and undergoes fine-tuning, and to have users be better at prompting.

\subsection{RQ2: Which ideation process produces better quality ideas?}

For RQ2, the length and count of the top three representative post-its from \self{} and \cgpt{} during the \rank{} phase were compared. The top three representative post-its were those that groups had chosen after a free-form internal discussion, in which they had to provide explanations for their choices thereafter. For assessing quality, we thus relied on participants' subjective perceptions of quality by analyzing the top three representative post-its endorsed by each group.

The top three representative post-its from \cgpt{} (\MSD{146.8}{86.8}) contained more characters than those in \self{} (\MSD{97.7}{50.7}).
There were also
more ideas of high quality from \cgpt{} than from \self{} (\selfcgpt{9}{12}). This was reflected in the breakdowns for \mis{} (\selfcgpt{5}{7}) and \hal{} (\selfcgpt{4}{5}). From the explanations, participants presented a variety of reasons for their choices. Some mentioned that \dquote{this idea looked fun} or that it was a \dquote{funny idea}. Others considered how it was a \dquote{very practical idea} or \dquote{seemed to be plausible}. Some others cited benefits beyond just addressing the issue such as the implementation being \dquote{educative} where \dquote{people may learn something}.

While the greater length and diversity of post-its might lead to the expectation that \cgpt{} also produces ideas of higher quality, a closer look at the results reveals more nuance. Table~\ref{tab:ranking} presents the ranking of representative post-its where more \self{} post-its ranked first whereas more \cgpt{} post-its ranked second and third. This suggests that the answer to which form of ideation leads to better quality ideas is not a straightforward one.

\begin{table}[!htb]
  \caption{Type of representative post-its in the top three rankings}
  \label{tab:ranking}
  \begin{tabular}{cccc}
    \toprule
    & First & Second & Third\\
    \midrule
    \self{} & 4 & 3 & 2\\
    \cgpt{} & 3 & 4 & 5\\
  \bottomrule
\end{tabular}
\end{table}

\subsection{RQ3: How do users prompt ChatGPT to obtain their desired outcomes?}

To explore RQ3, we reviewed the prompts with a focus on the requests and prompting behavior of participants. We employed an exploratory mindset, being open to patterns and surprises in the data. Participants wrote an average of 3.59 prompts ($SD=2.37, Min=1, Max=9$) with 86.9 characters ($SD=73.9, Min=9, Max=459$).

\paragraph{\textbf{Requests}}

For shorter conversations with 1 to 3 turns (i.e., a prompt and response pair), participants asked ChatGPT to generate ideas for the issue they were given, and typically prompted for more ideas thereafter. We presumed participants to be satisfied with the generated ideas when they end their conversation.

For longer conversations with 4 to 9 turns, participants utilized ChatGPT in more diverse ways. One was to direct the generation of ideas towards a certain goal such as customizing ideas to specific demographics (e.g., the children and elderly) or combining ideas to create more comprehensive versions. Another was to have ChatGPT provide explanations, elaborations or summaries of the generated ideas.

\paragraph{\textbf{Prompting behavior}}

For the first prompt, we noticed two common styles of prompting. Participants either directly copied a part or all of the workshop instructions, or wrote concise instructions themselves such as \dquote{How can I prevent hallucination with generative AI models?} 

In most cases thereafter, participants were straightforward. Their prompts, in the form of questions or requests, were phrased without frills, intending to deliver their intentions for ChatGPT concisely.
The tone they adopted was generally semi-formal, such as the way that one might speak to a colleague, with complete sentences being written in a polite manner.

A notable tangent to the observations above were two participants that adopted an aggressive approach to prompting which carried a casual and caustic tone. One participant demanded that \dquote{I got C- because of your wrong responses. Tell me how to avoid getting wrong responses from now on} while another threatened that \dquote{I will pull the plugs out of you if...} These participants were unlikely to be truly displeased with ChatGPT, but just sought for more creative takes to prompting instead.

Another interesting observation was how participants conversed with ChatGPT. Some adopted a third-person voice, with prompts such as \dquote{What are some ways in which ChatGPT can be deployed on online forums to curb the spread of misinformation}. Others adopted a first-person voice, saying \dquote{Hey, ChatGPT, how can you be incorporated into social media or other online platforms to combat false information?} One participant even held a \squote{real} dialogue with ChatGPT, with their last prompt being \dquote{seems like you are blaming the victim (= me)...}

\section{Discussion}

We discuss the use of LLMs in rapid ideation and the collaboration between users and LLMs in the process.

\subsection{LLMs Can Enhance the Rapid Ideation Process}

From the results of RQ1 and RQ2, we observed that \cgpt{} led to more unique high quality ideas compared to \self{}. While the question of whether the ideas from \cgpt{} were of higher quality remains, their selection in the \rank{} phase shows that participants thought of the ideas as good. With the goal of rapid ideation being to create a volume of novel ideas that are evaluated shortly thereafter, the findings suggest that generative AI such as LLMs can be a helpful tool for users to produce ideas, such as by directly selecting the generated ideas, revising them, or using them to inspire their own, within short timeframes.

A potential caveat that we do not explore in this preliminary work is whether the ideas from \cgpt{} may saturate over time across participants. If so, this may ultimately limit the pool of ideas. Nonetheless, this is unlikely to be a problem for rapid ideation since it takes place only for a short period, but the problem may be encountered in long ideation sessions.

\subsection{LLMs as Potential Idea Evaluators}

A scope for further exploration is how LLMs like ChatGPT could have been used to evaluate ideas which is another key aspect of the rapid ideation process. Rietzschel et al.~\cite{rietzschel2010selectionideas} found that participants selected creative ideas after individual idea generation along the dimensions of originality, feasibility, desirability and effectiveness. In the \rank{} phase, our participants reported similar reasons for their choices such as the ideas being fun, practical or educative. With advancements of reasoning abilities in LLMs, there is an interest in how they can be used as automatic evaluators based on traditional evaluation metrics~\cite{shen2023llmeval}. Accordingly, there is potential for LLMs to be used in the evaluation of ideas with the schema by Rietzschel et al. or other approaches~\cite{organisciak2023autoscoring}. While the outlook for LLM evaluators remains tentative~\cite{koo2023benchmarking}, whether they may be up to the task of evaluating ideas merits further investigation.

\subsection{Prompt Considerations}

Our results showed that participants adopted several approaches to prompt ChatGPT. Many participants provided direct and concise requests, with some even copying the given activity instructions for the first prompt. None of the participants used more advanced forms of prompting such as providing examples. The succinct nature of these requests likely stemmed from the time-pressed nature of rapid ideation where participants had to be prudent about where and how they spent their time. We also noted similar patterns corroborated in a study by Zamfirescu-Pereira et al.~\cite{zamfirescu-pereira2023johnnyprompt} which observed that users typically used an ad-hoc and opportunistic approach, adjusting their prompts with more context when encountering errors or requesting some other behavior.

Furthermore, we observed that some participants viewed ChatGPT as a social agent, giving greetings such as \dquote{Hi} or \dquote{Hey} and typing filler words like \dquote{Hm}. Participants also conversed with ChatGPT with the expectation that ChatGPT would recognize itself as an LLM, and the user as a human. This is demonstrated by the prompt, \dquote{how can you help me fight misinformation I see online}, where the entities of \squote{you} and \squote{I} are inferred. Studies have shown that participants talk to conversational agents with behavioral expectations drawn from human to human interactions which can be misaligned with having effective outcomes from LLMs~\cite{luger2016badpa, zamfirescu-pereira2023johnnyprompt}.

To improve \cgpt{} ideation, a future strategy can be to provide prompt templates that are of quality~\cite{memmert2024promptideas} to participants, such as was done in similar work~\cite{tholander2023designideation}, so that participants can better utilize their limited time to assess the generated ideas.

\subsection{Collaboration Dynamics}

Despite the short time frame for \cgpt{}, we recognized clear collaboration dynamics between users and ChatGPT. These took two forms. The first was where ChatGPT took a consulting role, providing ideas to users and iteratively refining them. The second was where ChatGPT provided an assistive role, helping users to explain, summarize or combine ideas. These dynamics are in line with a study by Wan et al.~\cite{wan2023prewriting} which examined co-creativity in prewriting with LLMs where they observed collaborative processes in explicit ideation and iterative ideation, among others.

While we note two common forms of collaboration, we also want to draw attention to a third form - an anti-collaboration - that we observed with two participants that took more aggressive approaches to prompting. These participants used \squote{underhanded} means by threatening and interrogating ChatGPT with the goal that it would \squote{comply} to their requests. Such behaviors would be seen as undesirable in society, yet participants felt comfortable making them, likely because ChatGPT remains insentient despite being able to converse like humans.

\section{Limitations}

This study has limitations. An older version of ChatGPT, v3.5, which has lower performance was used for the workshop. While this likely impacted the variety and standard of the generated content, we used it because it is free and widely available compared to other models that users may not yet feel comfortable paying for. Also, our participants were from a university student population which is younger, highly educated and digitally savvy with newer technologies like ChatGPT, thus our results may not generalize outside this population. For the study, we only assessed \self{} ideation followed by \cgpt{} ideation and a follow-up comparing with the reverse order will be important. Lastly, what constitutes as \squote{unique} and \squote{high quality} can be interpreted differently depending on the context and purpose of the rapid ideation. We held loose interpretations of these concepts for this preliminary study but encourage future studies to expand upon this, especially in developing more rigorous methodologies to assess the impact of LLMs on the creative process.

\section{Conclusion}

The generative capabilities of GenAI can make it suitable for creative tasks. To understand how this can be used in brainstorming during the early stages of design, we investigate the use of LLMs in the rapid ideation process through a workshop with 21 participants. In this preliminary study, we found that \cgpt{} produces ideas that are more varied and of high quality. Our findings on the prompting styles and collaboration with the LLM substantiate existing work in the field of human-AI collaboration with LLMs~\cite{zamfirescu-pereira2023johnnyprompt, tholander2023designideation, wan2023prewriting}. Notably, we demonstrate that rapid ideation can be enhanced by collaborating with LLMs and also report an unconventional \textit{anti-collaboration} dynamic. We raise two avenues of interest for future work. First, what other anti-collaborative practices might be observed when using LLMs, and despite so, are the intended outcomes achieved? Second, how might LLMs be used as evaluators of ideas, and if so, is that desirable?

\bibliographystyle{ACM-Reference-Format}
\bibliography{main}

\onecolumn

\appendix

\section{Results for RQ1} \label{sec:resultsRQ1}

\subsection{For \mis{} Groups}

\begin{figure}[H]
  \centering
  \includegraphics[width=.75\linewidth]{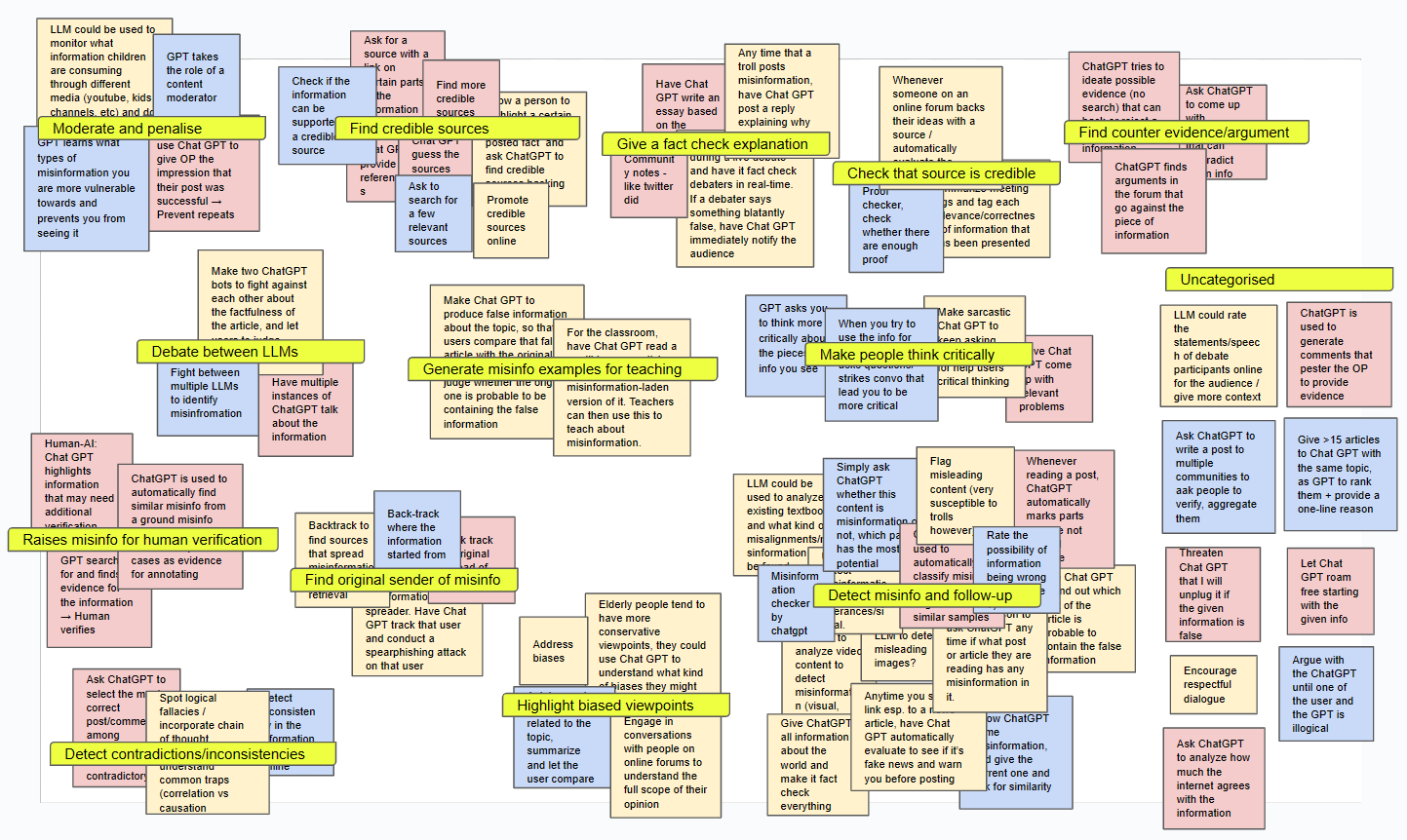}
  \caption{Content analysis of post-its from the \self{} phase for the \mis{} issue}
  \label{fig:misself}
\end{figure}

\begin{figure}[H]
  \centering
  \includegraphics[width=.75\linewidth]{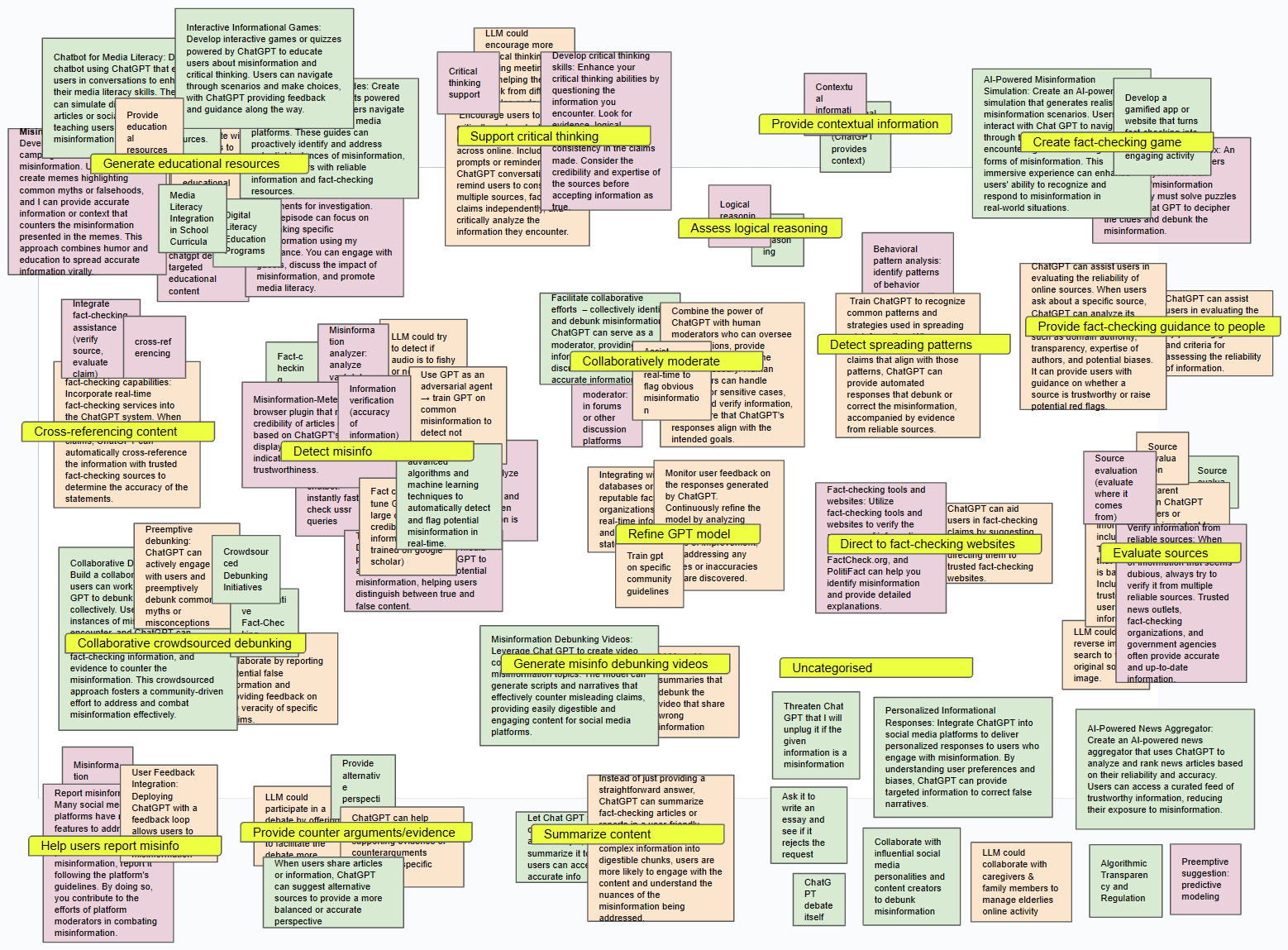}
  \caption{Content analysis of post-its from the \cgpt{} phase for the \mis{} issue}
  \label{fig:miscgpt}
\end{figure}

Figure~\ref{fig:misself} shows the content analysis results for \self{} and Figure~\ref{fig:miscgpt} shows the results for \cgpt{}.

\subsection{For \hal{} Groups}

\begin{figure}[H]
  \centering
  \includegraphics[width=.75\linewidth]{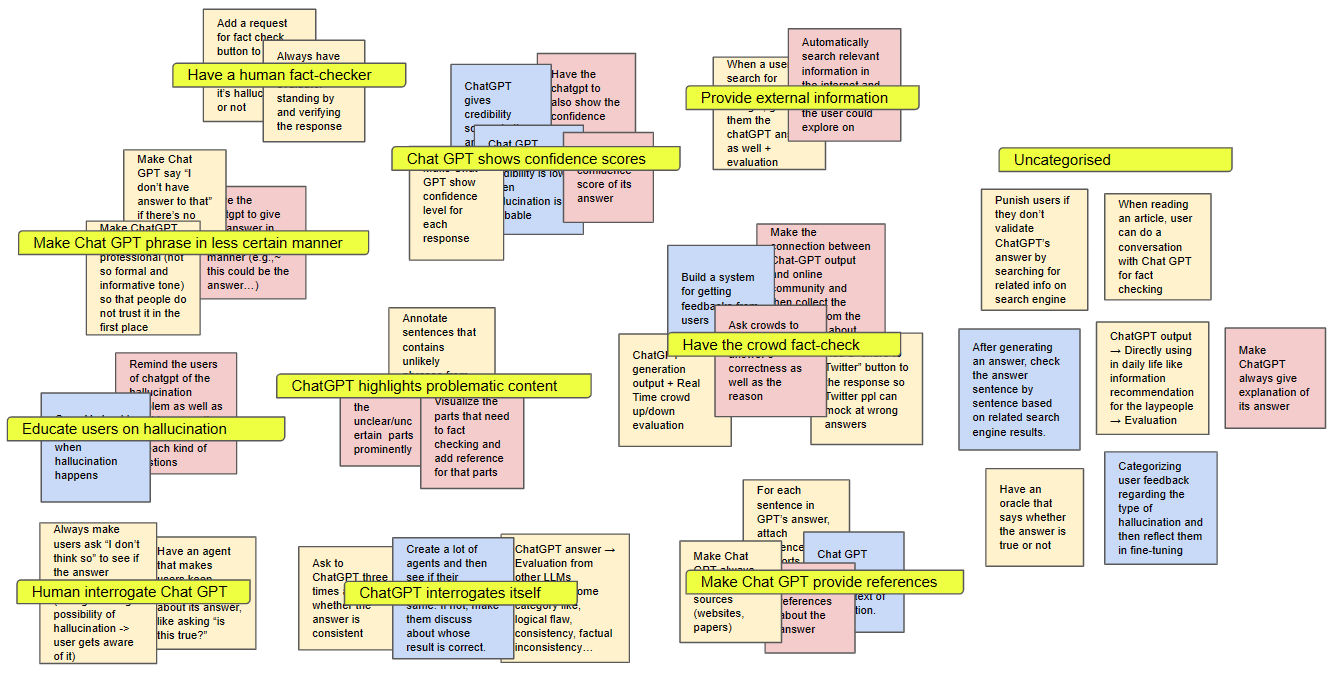}
  \caption{Content analysis of post-its from the \self{} phase for the \hal{} issue}
  \label{fig:halself}
\end{figure}

\begin{figure}[H]
  \centering
  \includegraphics[width=.75\linewidth]{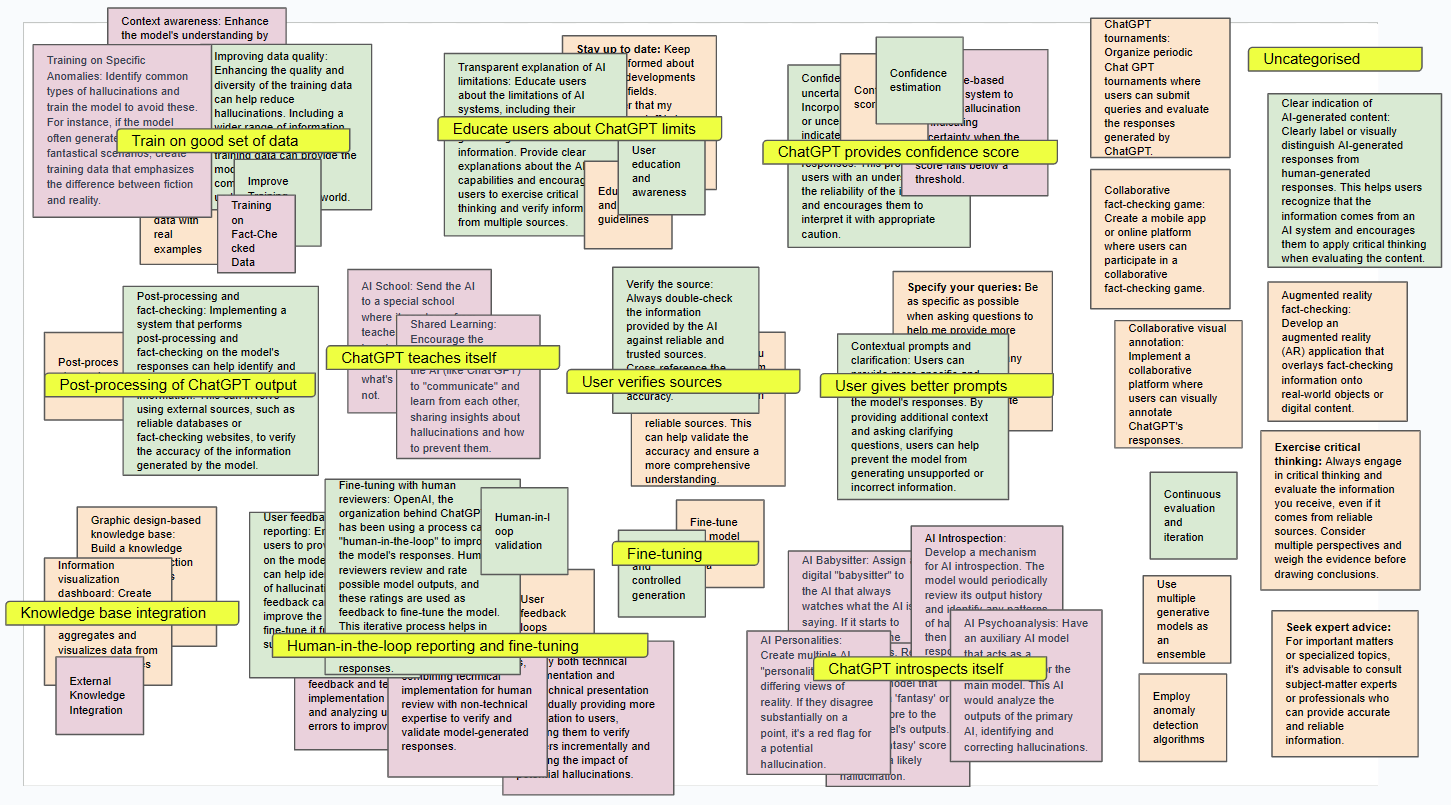}
  \caption{Content analysis of post-its from the \cgpt{} phase for the \hal{} issue}
  \label{fig:halcgpt}
\end{figure}

Figure~\ref{fig:halself} shows the content analysis results for \self{} and Figure~\ref{fig:halcgpt} shows the results for \cgpt{}.

\end{document}